\documentclass{emulateapj}
\shorttitle{Saturn as a Transiting Exoplanet}
\shortauthors{Dalba et al.}

\usepackage{subfigure}
\usepackage{url}
\usepackage{hyperref}
\usepackage{epsfig}
\usepackage{graphicx}
\usepackage{sidecap}
\usepackage{hanging}
\usepackage{color}
\usepackage{multirow}
\usepackage{enumerate}
\usepackage{natbib}
\usepackage{amssymb}
\usepackage{pdflscape}
\usepackage{amsmath}

\newcommand{\Cassini}{{\it Cassini} }
\newcommand{\Kepler}{{\it Kepler} }
\newcommand{\HST}{{\it HST} }
\newcommand{\Spitzer}{{\it Spitzer} }

\providecommand{\e}[1]{\ensuremath{\times 10^{#1}}}
\newcommand{\emcee}{\texttt{emcee }}

\begin{document}

\title{The Transit Transmission Spectrum of a Cold Gas Giant Planet}

\author{Paul A. Dalba,\altaffilmark{1} Philip S. Muirhead,\altaffilmark{1} Jonathan J. Fortney,\altaffilmark{2} Matthew M. Hedman,\altaffilmark{3} Philip D. Nicholson,\altaffilmark{4} Mark J. Veyette\altaffilmark{1}}

\affil{\vspace{0pt}\\ $^{1}$Department of Astronomy, Boston University, 725 Commonwealth Ave., Boston, MA, 02215, USA; \href{mailto:pdalba@bu.edu}{pdalba@bu.edu} \\
$^{2}$Department of Astronomy \& Astrophysics, University of California, Santa Cruz, CA 95064, USA \\
$^{3}$Department of Physics, University of Idaho, Moscow, ID 83843, USA \\
$^{4}$Department of Astronomy, Cornell University, Ithaca, NY 14853, USA}

\begin{abstract}

We use solar occultations observed by the Visual and Infrared Mapping Spectrometer aboard the \Cassini Spacecraft to extract the 1 to 5 $\mu$m transmission spectrum of Saturn, as if it were a transiting exoplanet. We detect absorption from methane, ethane, acetylene, aliphatic hydrocarbons, and possibly carbon monoxide with peak-to-peak features of up to 90 parts-per-million despite the presence of ammonia clouds. We also find that atmospheric refraction, as opposed to clouds or haze, determines the minimum altitude that could be probed during mid-transit. Self-consistent exoplanet atmosphere models show good agreement with Saturn's transmission spectrum but fail to reproduce a large absorption feature near 3.4 $\mu$m likely caused by gaseous ethane and a C-H stretching mode of an unknown aliphatic hydrocarbon. This large feature is located in one of the \textit{Spitzer Space Telescope} bandpasses and could alter interpretations of transmission spectra if not properly modeled. The large signal in Saturn's transmission spectrum suggests that transmission spectroscopy of cold, long-period gaseous exoplanets should be possible with current and future observatories. Motivated by these results, we briefly consider the feasibility of a survey to search for and characterize cold exoplanets analogous to Jupiter and Saturn using a target-of-opportunity approach.   

\end{abstract}

\keywords{planets and satellites: atmospheres --- planets and satellites: individual (Saturn) --- stars: planetary systems}
\maketitle

\section{Introduction}

To date, investigations of exoplanet atmospheres have not targeted those resembling the cold, gaseous planets in our solar system. Transit observations have only been used to characterize the atmospheres of exoplanets on short-period orbits \citep[e.g.][]{Charbonneau2002,Knutson2007,Pont2008,Kreidberg2014a,Fraine2014}. Close-in exoplanets are warm, producing favorable atmospheric scale heights, and they transit their hosts frequently, providing many opportunities to characterize their atmospheres. Direct imaging observations, which require that exoplanets be self-luminous, have also only been able to characterize warm exoplanets; most directly-imaged planets are younger than 50 Myr and hotter than 800 K \citep{Marois2008,Lagrange2010,Carson2013}. High-resolution, ground-based spectroscopy also favors short-period planets that experience large changes in radial-velocity during a single observation or have atmospheres warm enough to be observed in emission \citep[e.g.][]{Snellen2010,Birkby2013}. Other observing techniques (i.e. radial-velocity and microlensing) effectively discover long-period, giant exoplanets but do not provide information pertaining to planets' atmospheres.

Transmission spectroscopy is the most appropriate known method for characterizing the atmospheres of cold, long-period planets resembling those in our solar system. Unfortunately, transit surveys are geometrically biased against long-period planets \citep{Beatty2008}. According to the NASA Exoplanet Archive and the Exoplanet Orbit Database \citep{Han2014},\footnote{Both databases accessed 2015 July 23.} of the 1,230 confirmed exoplanets discovered via the transit method, only 5 (0.4\%) have orbital semi-major axes $a \gtrsim 1$ AU: Kepler-34b, Kepler-47c, Kepler-90h, Kepler-421b, and Kepler-452b \citep[][respectively]{Welsh2012,Orosz2012,Cabrera2014,Kipping2014a,Jenkins2015}. Of these, only Kepler-421b has an equilibrium temperature below 200 K (assuming a Bond albedo of 0.3), making it nearly as cold as Jupiter. Having only this one member, the regime of transiting exoplanets analogous to the giant planets in our solar system has so far gone relatively unexplored.  

Transmission spectra have the potential to reveal molecular abundances in exoplanet atmospheres, which constrain models of their thermal profiles \citep{Fortney2010}. Transmission spectra are also useful for full atmospheric retrieval codes \citep[e.g.][]{Madhusudhan2009,Irwin2008,Line2013,Line2014} that explore phenomena such as temperature inversions and disequilibrium chemistry, both of which have been observed in the solar system gas giants \citep{Baines2005,West2009,Bagenal2004,Tokunaga1983}.

Atmospheric abundances of molecules such as CO, CH$_4$, CO$_2$, and H$_2$O place constraints on C/O, C/H, and O/H ratios, which are tracers of planetary formation and evolution. Many planet formation theories, including that for Jupiter, invoke core accretion \citep{Owen1999,Pollack1996}, which has been tested by observations of C/O in hot Jupiters \citep[e.g.][]{Stevenson2014,Brogi2014,Line2014}. Atmospheric abundance measurements of cold, giant exoplanets would provide a similar test of core accretion and could also be used to improve the current understanding of how atmospheric abundances respond to planetary migration \citep{Ida2004,Madhusudhan2014,Oberg2011}.  

Atmospheric abundances can be difficult or impossible to determine for atmospheres that harbor clouds or haze, which produce flat transmission spectra across near infrared wavelengths \citep[e.g.][]{Bean2010,Berta2012,Kreidberg2014a,Knutson2014a,Knutson2014c,Ehrenreich2014}. Clouds are present in the atmospheres of each giant solar system planet \citep[e.g.][]{West2009,Bagenal2004,Lindal1987,Smith1989}, but these planets are much colder and experience different levels stellar insolation than previously-observed cloudy exoplanets. It is not clear how these differences would influence the effects of clouds on the transmission spectrum of a cold giant exoplanet.

To begin exploring the regime of cold, long-period exoplanets, we turn to an extensively studied gas giant in our own solar system: Saturn. High-quality solar system observations provide a unique opportunity to study and ``ground-truth'' the methods used to characterize exoplanets. Solar system bodies such as the Earth \citep[e.g.][]{VidalMadjar2010,GarciaMunoz2012,Betremieux2013,Misra2014,Schwieterman2015}, Titan \citep{Robinson2014}, Jupiter \citep{Irwin2014,MontanesRodriguez2015}, Uranus and Neptune \citep{Kane2011} have all been studied in the context of extrasolar planetary science in recent years.

We use observations from the Visual and Infrared Mapping Spectrometer \citep{Brown2004} aboard the \Cassini Spacecraft to extract the 1 to 5 $\mu$m transmission spectrum of Saturn, as if it were a transiting exoplanet. With this spectrum, we asses the feasibility of observing cold, gaseous exoplanets with current and future observatories. 

We present the \textit{Cassini}-VIMS observations, data, and analysis procedures in Section \ref{sec:obs}. In Section \ref{sec:model} we develop an occultation model that assumes the portion of Saturn's atmosphere sampled by the observations is isothermal and in hydrostatic equilibrium. We also fit for the effects of atmospheric refraction and absorption versus ascribing them from previous observations. In Section \ref{sec:results}, we calculate the transmission spectrum of Saturn and compare it to spectra of model atmospheres that are currently applied to exoplanets. Lastly, in Section \ref{sec:disc} we discuss the implications of this work for exoplanet atmosphere models, and we briefly consider a strategy to locate and characterize cold, giant exoplanets in the near future.

\section{Observations}\label{sec:obs}

\subsection{\textit{Cassini}-VIMS}

The Visual and Infrared Mapping Spectrometer (VIMS) aboard the \Cassini Spacecraft has been observing Saturn and its satellites since arriving at the Saturnian system in 2004 \citep{Matson2002,Brown2004}. VIMS has two imaging grating spectrometers, VIMS-VIS and VIMS-IR, that operate in the visible (0.35-1.07 $\mu$m, 96 bands, $\sim$8 nm resolution) and near-IR (0.85-5.11 $\mu$m, 256 bands, $\sim$17 nm resolution), respectively. Only the latter is used during solar occultations. Solar occultation observations are obtained through a solar port with an aperture of 30 mm by 5 mm that is orientated 20$^{\circ}$ away from the boresight direction of the main aperture. In the solar port, sunlight undergoes several reflections that attenuate the solar flux by a factor of approximately 2.5\e{-7} before passing through the slit and in to the main optical path of VIMS-IR \citep{Bellucci2009}. The nominal VIMS-IR observation produces a data cube comprised of two spacial dimensions (64 x 64 pixels) and one spectral dimension. The indium antimonide IR detector is a one-dimensional array (1 x 256 pixels), so it can only obtain the spectrum of a single spacial pixel at a time. Therefore, the IR telescope's secondary mirror is scanned in two dimensions across the target to construct a full data cube. For solar occultation observations, VIMS only acquires a 12 x 12 pixel field-of-view, which corresponds to an angular size of 20\farcm6 x 20\farcm6 (each pixel having an angular resolution of 1.7 arcminutes). This reduction in field-of-view is acceptable since the solar disk as seen from Saturn only extends over approximately 2 x 2 pixels.

\subsection{Occultation Data}

We analyzed a Saturn solar occultation observation from UT 2007 November 17. The observation consisted of 479 data cubes, each having an image dimension of 12 x 12 pixels and an exposure time of 20 ms per pixel. VIMS began observing several minutes before ingress in order to establish a high-signal measurement of the solar spectrum out of occultation. Similarly, the observation ended several minutes after the solar flux was completely attenuated. The duration of an entire observation was approximately 0.5 hours.

For each 12x12-pixel image, we determined a value of relative transmission ($T_{\lambda}$) by summing the signal over the entire field-of-view and dividing by the total signal of the Sun prior to occultation. Outside of occultation, $T_{\lambda} \sim 1$. Once the Sun's flux was completely attenuated by Saturn, $T_{\lambda} \sim 0$. We followed this normalization procedure for each of the 256 wavelength bands in each of the 479 data cubes. This ratio removed systematic and instrumental errors along with the the need to convert the detector's data numbers (or counts) into specific energies. Data calibration was further simplified by the high linearity of VIMS-IR detector \citep{Brown2004} and by the low background signal compared to that of the Sun (less than 1\%). The data we considered did not suffer from contamination by stray light entering the boresight, spacecraft pointing instability, or other sources of spurious signal that warrant advanced calibration procedures \citep[e.g.][]{Maltagliati2015}. A formal data reduction routine for VIMS exists \citep{McCord2004}; however, it is not appropriate for observations of solar occultations that pass through a different optics chain than those acquired through the main aperture. 

When calculating $T_{\lambda}$, we defined the background as all the pixels residing outside of a circular aperture centered on the Sun with a radius of four pixels. The average background was approximately 14 counts per pixel, only 0.6\% of the average integrated signal from the Sun ($\sim$2,260 counts). Since the background level decreased unevenly as Saturn's limb entered the field-of-view, we could not accurately estimate the background signal across the detector simply by finding the mean number of counts in a circular annulus surrounding the central aperture. Instead, we separated the detector into four 6x6-pixel quadrants, and subtracted the mean background locally in each. When Saturn's atmosphere was in view, the average background value was 1-2 counts per pixel. Considering the minimal contribution of the background to the total count value of the entire field-of-view, this simple procedure was sufficient. 

After calculating $T_{\lambda}$, we median-filtered each occultation light curve to remove outliers due to other sources of spurious signal, most which were cosmic ray strikes on the detector. A data point was declared an outlier and removed if it had a value of $T_{\lambda}$ that was either 3$\sigma$ above or below the median $T_{\lambda}$ value of the six points on either side of it. 

We assigned each $T_{\lambda}$-value an uncertainty equal to the standard deviation of solar signal prior to occultation. At redder wavelengths ($\lambda>$ 4 $\mu$m), the solar intensity was weak and the data became increasingly noisy. The reddest 8 bands spanning 4.99--5.12 $\mu$m were used to record timing information for the observations and were not included in the following analysis. Some of the VIMS data exhibited low-level, time-correlated noise, possibly due to detector readout effects \citep{McCord2004}. Its magnitude was typically on the order of the uncertainty and did not greatly affect the signal or the analysis. 

\begin{figure}
\centering
\includegraphics[scale=0.44]{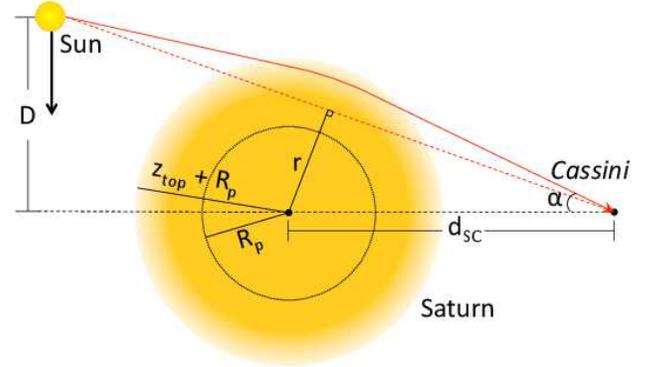}
\caption{Geometry of the Sun-Saturn-\Cassini system (not to scale, rings of Saturn not pictured). Light from the Sun followed a curved path in Saturn's atmosphere (solid red line). The tangent radius ($r$) was measured from the center of Saturn to the point along the straight line-of-sight between \Cassini and the Sun (dashed red line) that was tangent to the local horizon of Saturn. In the model, rays from the Sun entered Saturn's atmosphere at an altitude of $R_{\rm p} + z_{\rm top}$ before reaching \Cassini with angle $\alpha$ to the \textit{Cassini}-Saturn line. $R_{\rm p}$ was the ``surface'' of Saturn from which the altitude $z$ was measured. As the occultation progressed, the Sun appeared to move in the direction indicated by the black arrow from the point of view of \textit{Cassini}. Each value of $r$ corresponded to a value of $D$ (\S\ref{sec:D}). There was some impact parameter between the path of the Sun and the center of Saturn as seen from \Cassini (i.e. the Sun did not pass directly behind the center of Saturn). However, the occultation model only tracked the one-dimensional radial motion of the Sun.}
\label{fig:geom}
\end{figure}

We monitored the progress of the occultation with measurements of the ``tangent radius'' $r$. This was a measure of distance between the center of Saturn and the point on a straight line-of-sight between the Sun and \Cassini that was tangent to the local horizon of Saturn (see Fig. \ref{fig:geom}). We used $r$ as a substitute for time since it included information about the relative positions of the Sun, Saturn, and \Cassini that was useful when modeling the occultation. 

\section{A Solar Occultation Model}\label{sec:model}

\subsection{Parametrizing Saturn's Atmosphere}

A goal of this work was to measure Saturn's transmission spectrum as empirically as possible. Therefore, we modeled the Saturn-solar occultations without directly using atmospheric chemical abundances, mixing ratios, indices of refraction, and opacities available in the literature. Instead, we fit the VIMS occultation data to a model atmosphere and estimated parameters describing the structure and composition of Saturn's atmosphere. Each of \textit{Cassini}-VIMS' wavelength bands had its own best-fit occultation light curve. For each wavelength band, we assumed the portion of Saturn's atmosphere sampled by the observation was ideal, isothermal, and in hydrostatic equilibrium in order to acquire the familiar number density profile

\begin{equation}\label{eq-nd}
n(z) = n_0 e^{-z/H}
\end{equation}

\noindent where $z$ is altitude, $n(z)$ is particle number density as function of altitude, and $n_0$ is the reference particle number density at the $z=$ 0 m ``surface'' of Saturn ($R_p$), which approximately corresponded to the one-bar pressure level. $H$ is the atmospheric scale height defined by

\begin{equation}\label{eq:scale}
H = \frac{k_B \mathcal{T}}{\mu m_p g}
\end{equation}

\noindent where $k_B$ is Boltzmann's constant, $\mathcal{T}$ is temperature, $\mu$ is the mean molecular weight of Saturn's atmosphere, $m_p$ is the proton mass, and $g$ is the local acceleration due to gravity. $H$ was a critical parameter to the occultation model as it controlled how steeply the transmission decreased during ingress. The scale height did not have a wavelength dependence \textit{per se}, but we could not use a single value of $H$ in the model across the spectrum. Due to methane absorption, different wavelengths sampled portions of Saturn's atmosphere that were separated by up to $\sim$450 km in altitude. This was readily observable in the occultation data as a range in ``half-light'' $r$-values, where $T_{\lambda}=$ 0.5. Over $\sim$450 km, variations in temperature and therefore scale height necessitated that we fit for $H$ at each wavelength in the model.

We used two parameters to describe the wavelength-dependent absorption and refraction of light in Saturn's atmosphere: the {\it total} absorption cross section $\sigma_{\lambda}$ and the {\it total} refractivity\footnote{The refractivity ($\nu$) is related to the index of refraction ($\eta$) by $\nu=$ $\eta$ - 1.} $\nu_{\lambda}$. Both parameters included contributions from all atmospheric species. The other parameters in the model, $R_p$ and $n_0$, were not wavelength-dependent so we adopted the following one-bar values for Saturn: $R_p=$ 5.7\e{7} m and $n_0=$ 5.5\e{25} m$^{-3}$ \citep{Hubbard2009,Fouchet2009,West2009}. This value of $R_p$ accounted for Saturn's oblateness ($\sim$0.0979) and the local Saturn-centric latitude of observation ($\sim$49$^{\circ}$S), assuming Saturn to be an oblate spheroid \citep[e.g.][]{Smith1983,Cox2000}.

In a vertically stratified atmosphere, refractivity $\nu (z)$ can be defined as

\begin{equation}\label{eq:sum}
\nu (z) = \left [ \frac{n(z)}{L_0} \right ] \sum_j f_j (z) \nu_{0,j}
\end{equation}

\noindent where $L_0$ is Loschmidt's Number, $f_j(z)$ is the altitude-dependent mole fraction of the $j$th atmospheric species, and $\nu_{0,j}$ is the refractivity of the $j$th species at standard temperature and pressure. Loschmidt's Number is merely a particle number density at some standard temperature and pressure, both of which are functions of altitude in Saturn's atmosphere. We set $L_0 = n_0$, the reference particle number density, such that $n(z)$/$L_0=$ $e^{-z/H}$. We also assumed that Saturn's atmosphere was well-mixed such that $f_j$ did not have a $z$-dependence. This allowed us to treat the summation term in Eq. \ref{eq:sum} as a parameter and rewrite the entire equation as

\begin{equation}\label{eq:nu_z}
\nu (z) = \nu_{\lambda} e^{-z/H}
\end{equation}

\noindent where $\nu_{\lambda}$ is the wavelength-dependent total refractivity parameter described above.

We note that $\nu_{\lambda}$ was evaluated at $z=0$ m allowing for $\nu (z)$ to be calculated elsewhere in the atmosphere with Eq. \ref{eq:nu_z}. For $\sigma_{\lambda}$, we assumed a well-mixed composition at the altitudes sampled by the observation at a given wavelength so that $\sigma_{\lambda}$ did not have a $z$-dependence.

\subsection{Ray Tracing}

We traced rays between the Sun and \Cassini according to a ray tracing scheme developed by \citet{Kivalov2007}.\footnote{A concise summary of this ray tracing scheme was provided by \citet{vanderwerf2008}.} Each ray had finite energy and could be bent by refraction and attenuated by absorption. The density of rays in a given area and solid angle represented the specific intensity from the Sun.

At the time of observation, we determined the orbital distance of Saturn ($a=$ 9.524 AU) and the distance between \Cassini and the center of Saturn ($d_{\rm SC}=$ 2.59\e{8} m) using the JPL-HORIZONS solar system ephemeris computation service \citep{Giorgini1996}. We assumed these distances were constant during the 0.5-hour observation period.

We considered rays that reached the spacecraft at a positive angle of $\alpha$ relative to the {\it Cassini}-Saturn line that ranged from zero to $\arcsin{[(R_p+z_{\rm top}) / d_{\rm SC}]}$ where $z_{\rm top}$ was the fiducial ``top'' of Saturn's atmosphere equal to 1.2\e{6} m or $\sim$20 scale heights (see Fig. \ref{fig:geom}). At this altitude, the particle number density was reduced by a factor of 2\e{-9} from $n_0$ and the atmosphere was essentially transparent.

The ray tracing scheme accounted for refraction by modeling each step of a ray's motion through Saturn's atmosphere as a circle segment where the radius of curvature was a function of the index of refraction \citep{Kivalov2007,vanderwerf2008}. At each step, we calculated the optical depth ($\tau_{\lambda}$) experienced by the ray according to 

\begin{equation}\label{eq:tau}
\frac{\mathrm{d}\tau_{\lambda}}{\mathrm{d}s} = n(z) \sigma_{\lambda}
\end{equation}

\noindent where $s$ is the ray path length. The rays propagated through Saturn's atmosphere until one of two conditions was met: 1) $z=z_{\rm top}$ meaning that the ray reached the edge of the atmosphere on the \textit{Cassini}-side, or 2) $\tau_{\lambda}=$ 50 in which case the ray's energy had been fully attenuated.

\subsection{Generating a Transmission Model}\label{sec:D}

We made the occultation model in a reference frame such that Saturn and \Cassini were fixed relative to each other and the Sun appeared to move\footnote{This choice increased the computational efficiency of the model-fitting process.} in a plane perpendicular to the {\it Cassini}-Saturn line (see the black arrow in Fig. \ref{fig:geom}). This plane will herein be referred to as \textit{the plane of the Sun}. Positions on this plane with respect to the {\it Cassini}-Saturn line were expressed with the coordinate $D$. Although $D$ did not have a physical meaning, it allowed for direct comparison between the position of the Sun (from the data) and the rays' points of origin (from the model). The $D$-values of the Sun were calculated by projecting $r$, the tangent radius, back to the plane of the Sun using the geometry of the system.

\begin{figure}
\centering
\includegraphics[scale=0.36]{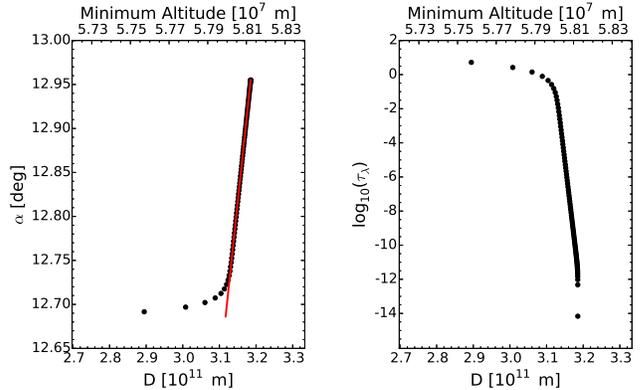}
\caption{Numerical relations between $\alpha$, $\tau_{\lambda}$, $D$, and minimum altitude at 1.25-$\mu$m in the  occultation model. Each ray (black data point) originated at the Sun with a $D$-value that corresponded to a value of $\alpha$ and $\tau_{\lambda}$, which governed brightness losses by refraction and absorption, respectively. The minimum altitude in Saturn's atmosphere ($R_{\rm p} + z$) achieved by each ray is also shown. These smooth functions allowed for interpolation of any $D$-value. The red line in the left panel shows the relation $\tan{\alpha} = D/(a+d_{\rm SC})$ where $D$ and $(a+d_{\rm SC})$ are the opposite and adjacent sides of a right triangle, respectively, from Fig. \ref{fig:geom}. Rays that only traverse Saturn's upper atmosphere (large $\alpha$) lie along this red line because they do not experience high indices of refraction and therefore travel in nearly straight lines. However, rays deviate from the red line at lower $\alpha$ as refractive bending becomes more significant.}
\label{fig:a-t_vs_D}
\end{figure}

Each ray considered by the model could be described by three quantities: $\tau_{\lambda}$, the final optical depth the ray achieved upon exiting Saturn's atmosphere; $\alpha$, the angle above the {\it Cassini}-Saturn line at which the ray reached \textit{Cassini}; and $D$, the height on the plane of the Sun above the {\it Cassini}-Saturn line where the ray originated. Both $\tau_{\lambda}$ and $\alpha$ were important in determining the decrease in brightness during the occultation. Figure \ref{fig:a-t_vs_D} shows that these quantities had smooth, numerical relations amenable to interpolation. For any $D$-values occupied by the Sun during the occultation, we could numerically determine the $\tau_{\lambda}$ and $\alpha$ values of the Sun's rays. We also measured the minimum radial distance from the center of Saturn achieved by each ray. This distance was important in assessing the effects of refraction in the data (\S\ref{sec:occ_to_tran}) and was physically more informative than $D$.

\begin{figure}
\centering
\includegraphics[scale=0.37]{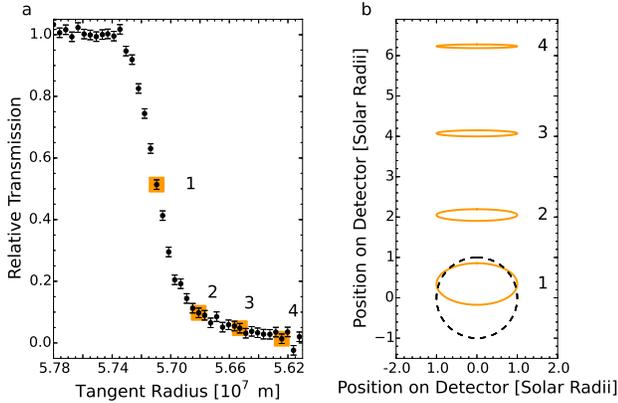}
\caption{Two effects of atmospheric refraction captured by the occultation model. \textbf{Panel a}: \textit{Cassini}-VIMS data (black data points) at 1.25 $\mu$m illustrating the decrease in transmission as the occultation progressed. \textbf{Panel b}: The apparent shape and position of the Sun on the VIMS detector predicted by the occultation model. The four numbered ellipses correspond to the four boxed data points in Panel a. The dashed circle is the shape and position of the unocculted Sun for reference. Since \Cassini always pointed towards the true position of the Sun, refraction caused the apparent position of the Sun to move against the gradient in $\nu$ (radially away from the center of Saturn) as the occultation progressed. The refractive spreading of the Sun's rays flattened the appearance of the solar disk into an ellipse. Each of these effects was present in the raw VIMS data cubes, although we did not use the image of the solar disk in the raw data cubes to estimate the parameters $\sigma_{\lambda}$ and $\nu_{\lambda}$. We display these phenomena simply to demonstrate that the occultation model correctly accounted for the effects of refraction.}
\label{fig:ref_effects}
\end{figure}

The $\tau_{\lambda}$-values allowed us to determine the energy attenuation due to absorption. The $\alpha$-values allowed us to determine flux losses due to refraction. Atmospheric refraction caused an apparent shrinking of the solar disk in the vertical (or radial) direction (see Fig. \ref{fig:ref_effects}). This change in shape resulted from the differential refraction experienced by rays originating at different points on the Sun. A ray leaving the ``top'' of the Sun traveled through a less dense portion of the atmosphere than a ray leaving from the ``bottom'' of the Sun. Consequently, the difference in $\alpha$ for these two rays and therefore the apparent angular size of the Sun on the detector diminished as the occultation progressed --- resulting in a loss of brightness. Another result of atmospheric refraction was the separation of the \emph{apparent} position of the Sun and the \emph{true} position of the Sun. Since \Cassini pointed towards the true position of the Sun throughout the occultation observation, this phenomenon manifested itself as an apparent motion of the Sun on the detector. Figure \ref{fig:ref_effects} illustrates that the occultation model accounted for both the apparent motion and shrinking of the solar disk.

\begin{figure}
\centering
\includegraphics[scale=0.42]{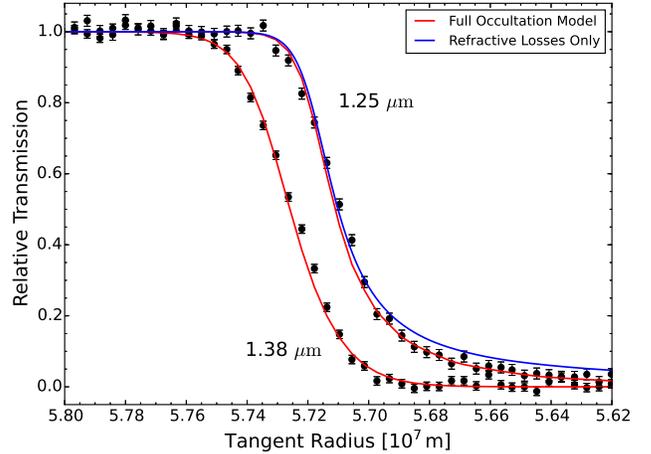}
\caption{\textit{Cassini}-VIMS data (black data points) and occultation model fits at 1.25 $\mu$m --- where CH$_4$ was transparent --- and 1.38 $\mu$m --- where CH$_4$ was opaque. Note that the tangent radius increases to the left. The dominant extinction process (refraction or absorption) and the shape of the transmission curves in these two wavelength channels were different. At 1.25 $\mu$m, the flux loss was almost entirely due to refraction, as shown by the blue curve which was found by ignoring absorption. At 1.38 $\mu$m, CH$_4$ absorption attenuated the solar flux before refractive loses became significant.}
\label{fig:model_fits}
\end{figure}

Having numerical functions for $\tau_{\lambda}(D)$ and $\alpha(D)$ meant that we could determine the relative transmission of the flux from any point on the solar disk throughout the entire occultation.\footnote{From the point-of-view of \Cassini, the Sun subtended 130 km in Saturn's atmosphere. At the altitudes considered in this work, the horizon of Saturn was virtually flat over 130 km. Therefore, we assumed Saturn's atmosphere was plane-parallel in calculating $T_{\lambda}$ over the entire solar disk.} By integrating over the solar disk, we calculated a model value of $T_{\lambda}$ for each $r$ and therefore a full occultation model. Examples of the \Cassini data and model fits in two characteristic wavelength bands are shown in Fig. \ref{fig:model_fits}.

\subsection{Bayesian Parameter Estimation}

We fit the occultation model to the data and extracted the best-fit values of $\sigma_{\lambda}$, $\nu_{\lambda}$, and $H$ using \texttt{emcee}, an open source, pure-\texttt{Python} Markov Chain Monte Carlo ensemble sampler \citep{ForemanMackey2013}. In each of the 248 wavelength bands, we applied uniform priors to $\sigma_{\lambda}$ and $H$ that restricted the parameter space to 1\e{-34} m$^2$ $ < \sigma_{\lambda} < $ 1\e{-29} m$^2$ and 2.0\e{4} m $< H <$ 8.0\e{4} m; any values of $\sigma_{\lambda}$ or $H$ outside of these ranges were considered to be unphysical based on our prior knowledge of Saturn's atmosphere. For $\nu_{\lambda}$, we imposed a normal prior with mean 2.5\e{-4} and variance 1.6\e{-9}. We chose these values based on known values of refractivity for H and He at a solar mixing ratio scaled to $\sim$ 134 K \citep[][]{Atreya1986} and based on the likely range of temperatures sampled by the occultation observations.    

\begin{figure}[!ht]
\centering
\includegraphics[scale=0.37]{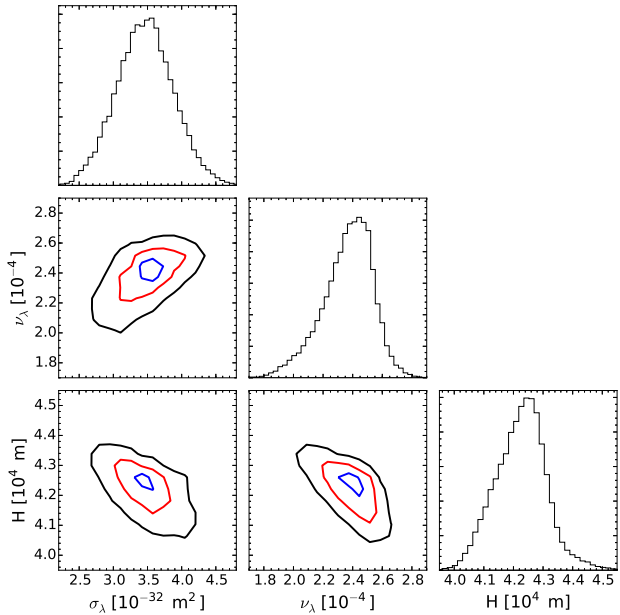}
\caption{Posterior probability distributions for the total absorption cross section ($\sigma_{\lambda}$), the total refractivity at one bar ($\nu_{\lambda}$), and the scale height ($H$) for a single wavelength band (1.25 $\mu$m). The one-dimensional histograms show the distributions for each parameter marginalized over the others and the two-dimensional histograms (with contours encompassing the 16th, 50th, and 85th percentiles) show the joint distributions for each parameter pair. The best-fit values and uncertainties found with these percentiles for $\sigma_{\lambda}$, $\nu_{\lambda}$, and $H$ at 1.25 $\mu$m were 3.47$^{+0.42}_{-0.42}$\e{-32} m$^2$, 2.38$^{+0.14}_{-0.18}$\e{-4}, and 4.229$^{+0.075}_{-0.094}$\e{4} m, respectively.}
\label{fig:corner}
\end{figure}

The posterior probability distribution function for the model parameters at 1.25 $\mu$m is shown in Fig. \ref{fig:corner}. The parameter distributions for most wavelengths were well-defined and Gaussian. In certain cases (e.g. 1.25 $\mu$m) the distributions for were skewed towards lower values of $\nu_{\lambda}$ and $H$. We noticed a slight correlation between these two parameters; the effect of increasing the scale height could be negated if $\nu_{\lambda}$ was allowed to reach unrealistic values greater than 1\e{-3}. Therefore, it was necessary to impose the aforementioned prior on refractivity. Parameter variances were higher in wavelength bands that exhibited higher noise and in two cases (1.64 and 3.88 $\mu$m) \emcee could not produce a well-defined posterior distribution. These wavelengths corresponded to two VIMS ``filter gaps'' where the spectral profiles of the channels were distorted \citep{Brown2004}. We did not include these channels in our calculation of Saturn's transmission spectrum. 

\subsection{Transforming from Occultation to Transit}\label{sec:occ_to_tran}

\begin{figure}
\centering
\includegraphics[scale=0.47]{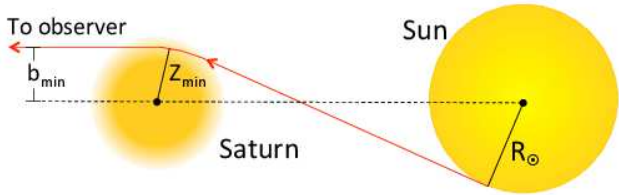}
\caption{Geometry of a Saturn-Sun exoplanet system at mid-transit (not to scale, rings of Saturn not pictured). The path of a maximally-deflected ray is shown in red. At mid-transit in regions of the spectrum where methane was transparent, atmospheric refraction determined the minimum altitude rays could probe ($z_{\rm min}$). Each $z_{\rm min}$ corresponded to a minimum impact parameter ($b_{\rm min}$) that set the continuum level of Saturn's transmission spectrum.}
\label{fig:transit}
\end{figure}

The occultation model returned parameters $\sigma_{\lambda}$ and $\nu_{\lambda}$ that described the opacity and refractivity of Saturn's atmosphere between 1 and 5 $\mu$m. With these parameters, we shifted from an occultation geometry, where the observer (\textit{Cassini}) was close to Saturn and relatively far from the Sun, to a transit geometry, where the observer was located at an infinitely large distance away from a Saturn-twin exoplanet orbiting a solar-twin star (see Fig. \ref{fig:transit}). In the transit geometry, the observer only measured rays that left Saturn's atmosphere parallel to the line-of-sight. These rays had a range of impact parameters ($b$) relative to the center of the exoplanet. While in the atmosphere, the rays still refracted according to Eq. \ref{eq:nu_z} and experienced attenuation according to Eq. \ref{eq:tau}, but the refractive spreading of the rays did not cause the apparent shrinking of the stellar disk that was present in the occultation observations. 

We considered the Saturn-Sun exoplanet system at the moment of mid-transit (Fig. \ref{fig:transit}). We traced rays with $\sim$4-km vertical resolution in the upper 3\% of Saturn's atmosphere\footnote{Rays sampling lower altitudes were absorbed.} to determine relations between impact parameter, final optical depth ($\tau_{\lambda}$), minimum altitude ($z_{\rm min}$), and point of origin on the Sun. A ray was considered to be absorbed if it reached $\tau_{\lambda} \ge 50$. We calculated the relative transmission of each ray using $T_{\lambda} = e^{-\tau_{\lambda}}$. The 4-km vertical resolution yielded smooth numerical relations between each of the above parameters allowing us to determine the transmission as a function of impact parameter, $T_{\lambda}(b)$, for Saturn at mid-transit. 

We also calculated the minimum impact parameter, $b_{\rm min}$, at each wavelength. In regions of the spectrum with high methane opacity, $b_{\min}$ corresponded to rays with final optical depths of $\sim$50. This meant that absorption limited the altitudes probed by the rays. Alternatively, in regions of the spectrum where methane was transparent, atmospheric refraction determined the value of $b_{\min}$ and the rays corresponding to $b_{\min}$ had optical depths less than unity. The significance of this result will be discussed in \S\ref{sec:no_ref}. 

We calculated the wavelength-dependent effective area of Saturn's disk ($A_{\rm eff, \lambda}$) using the expression

\begin{equation}\label{eq:area}
A_{\rm eff, \lambda} =\pi \left [ (R_{\rm p} + z_{\rm top})^2 - 2 \int_{b_{\rm min}}^{R_{\rm p} + z_{\rm top}} T_{\lambda}(b) \, b \, \mathrm{d}b \right ] \; ,
\end{equation}

\noindent which neglects the effects of stellar limb darkening \citep{Betremieux2014,Betremieux2015a}. The integral term subtracts circular annuli of thickness $db$ weighted by their relative transmission $T_{\lambda}(b)$ from the total combined area of the atmosphere and planet $\pi (R_{\rm p} + z_{\rm top})^2$. We then determined the value of transit depth $\delta_{\lambda}$ trivially using 

\begin{equation}
\delta_{\lambda} = \frac{A_{\rm eff, \lambda}}{\pi R_{\odot}^2} 
\end{equation}

\noindent where $R_{\odot}$ is the solar radius (6.96\e{8} m). The resulting transmission spectrum of Saturn is displayed in Fig. \ref{fig:tts_transit_all}.

We note that our method of removing the refractive flux losses intrinsic to occultation observations but not transit observations differed from the methods of \citet{Robinson2014}, who used \Cassini observations to measure the transit transmission spectrum of Titan. Instead of modeling Titan's atmosphere so that $T_{\lambda}(b)$ could be calculated in the case of a Titan-Sun exoplanetary system, \citet{Robinson2014} divided the \Cassini data by the correction factor 

\begin{equation}
f_{\rm ref} = \left ( 1+d_{\rm sc} \, \mathrm{d}\theta/\mathrm{d}z_{\rm min} \right )^{-1} \;, 
\end{equation}

\noindent where $\theta$ is the bending angle swept out by a ray due to atmospheric refraction. This factor is simply the occultation light curve that would be produced for the case of a completely transparent atmosphere such that brightness loss is \emph{only} due to refraction. The expression for $f_{\rm ref}$ was originally derived by \citet{Baum1953} under the assumption that $\theta$ was small or, equivalently, the index of the refraction was approximately unity \citep{Baum1953,Wasserman1973}.

As a sanity check, we recalculated the transmission spectrum of Saturn using the methods of \citet{Robinson2014}. The resulting transmission spectrum closely matched the one produced using the methods described in this work.

\section{Results}\label{sec:results}

\subsection{The Transmission Spectrum of Saturn}\label{sec:tts}

\begin{figure*}
\centering
\includegraphics[scale=0.59]{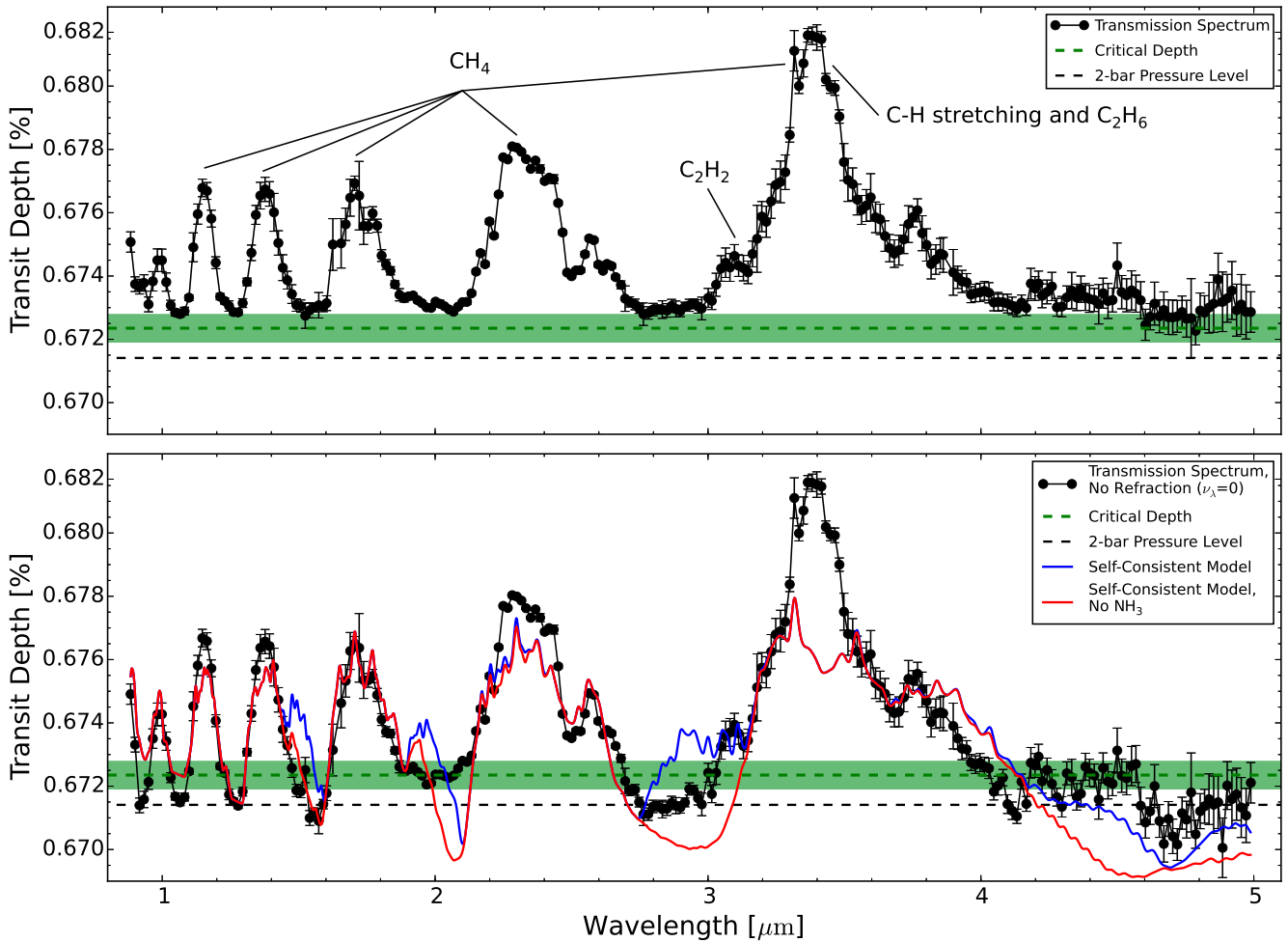}
\caption{\textbf{Top:} The near-infrared, transmission spectrum of Saturn (black data points). The error bars are the 1$\sigma$ uncertainties, which in some cases are smaller than the data point. The dashed green line and shaded green region correspond to the critical altitude and 1$\sigma$ uncertainty range, below which rays cannot probe during mid-transit due to atmospheric refraction. The dashed black line corresponds to the two-bar pressure level in the models and presumably the top of Saturn's global NH$_3$ cloud layer. \textbf{Bottom:} Saturn's transmission spectrum generated without the effects of refraction (see \S\ref{sec:no_ref}). In this scenario, the base of the spectrum is set by a gray opacity source near the two-bar level and not the critical depth. Two self-consistent atmosphere models (blue and red) are plotted with the transmission spectrum. The blue model allows for NH$_3$ in gaseous form while the red model forces the gaseous NH$_3$ content to zero. These models do not include the critical altitude set by refraction or the gray opacity source near two bars.}
\label{fig:tts_transit_all}
\end{figure*}

We generated the near-infrared transmission spectrum of Saturn as if it were a transiting exoplanet (Fig. \ref{fig:tts_transit_all}). The spectrum displays several spikes in transit depth of order 10 to 90 parts-per-million (ppm) corresponding to opacity from methane, ethane, acetylene, and possibly carbon monoxide between 4.1 and 5.0 $\mu$m. The largest feature, near 3.4 $\mu$m, is thought to be due to an asymmetric stretching mode of a C-H bond in an unknown aliphatic hydrocarbon chain. Similar chains have been identified in observations of Titan \citep{Bellucci2009,Robinson2014} and the diffuse interstellar medium \citep{Sandford1991}. A recent analysis of Titan solar occultations by \citet{Maltagliati2015} suggested that gaseous ethane may also contribute to the opacity between 3.2 and 3.5 $\mu$m. Gaseous ethane is present in Saturn's atmosphere \citep{Fouchet2009} and could therefore be contributing to the absorption near 3.4 $\mu$m. 

The uncertainties in Saturn's transmission spectrum are the standard deviations of 1,000 different transmission spectra, each calculated using different values for parameters $\sigma_{\lambda}$, $\nu_{\lambda}$, and $H$. The 1,000 different parameter sets formed a Gaussian distribution centered on the best-fit parameters values and with standard deviations equal to the uncertainties returned by \texttt{emcee}. The uncertainty was higher in the 4- to 5-$\mu$m region where the solar intensity was relatively weak.

While most of the features in the transmission spectrum were due to absorption, the baseline was determined by atmospheric refraction. This ``critical transit depth'' corresponded to a critical minimum altitude in Saturn's atmosphere that rays could probe during mid-transit. We found that the pressure level associated with the critical depth was 1.0 $\pm$ 0.5 bars. This value was consistent with a recent theoretical calculation of the critical pressure level for a Jupiter-sized planet with a 300-K atmosphere \citep{Betremieux2015b}. We note that we did not force this baseline; it is a simple geometric result of atmospheric refraction combined with the planet-star distance and the stellar radius. The baseline of the spectrum was located above a gray opacity source near two bars, which was presumably the top NH$_3$ cloud deck. As a result, signatures of this feature were not detected in Saturn's transmission spectrum. The value of the critical transit depth varied slightly across the spectrum due to the uncertainty in $\nu_{\lambda}$ and the minor wavelength-dependence of refractivity in Saturn's atmosphere.

\subsection{Self-Consistent Atmosphere Models}

Having ``reconstructed'' the transmission spectrum of Saturn using \textit{Cassini}-VIMS, we next calculated the transmission spectrum of a self-consistent ``off-the-shelf" atmosphere model for Saturn, following \citet{Fortney2005} and \citet{Fortney2010}. The philosophy was \emph{not} to search for a best fit, but rather to test how a model that was not tuned would fit the observations. As a transiting exoplanet, Saturn's surface gravity \citep[10.4 m s$^{-2}$,][]{Seidelmann2007} would be constrained, and the incident stellar flux around its G2V parent star at 9.524 AU \citep[from JPL-HORIZONS,][]{Giorgini1996} would be known. Furthermore, from the stellar age and planet mass the intrinsic flux from the planet's interior could be assessed from evolution models \citep[i.e.][]{Fortney2007}. With these parameters, the atmosphere code found a solution for the pressure-temperature-abundance profile that was in radiative-convective equilibrium given our knowledge of equilibrium chemistry and the wavelength-dependent opacity of each molecule. The code excluded photochemistry. Rather than solar abundances, we chose a metal-enhanced chemistry grid at 10$\times$ solar, as suggested from solar system and exoplanet trends \citep{Kreidberg2014b}. The transmission spectrum of the model was calculated using the one-dimensional code described in \citet{Fortney2010}.

The transmission spectrum of the self-consistent atmosphere model is shown in the bottom panel of Fig. \ref{fig:tts_transit_all}. Since the self-consistent models did not not include the limiting effects of refraction or the gray opacity source near two bars, it was more appropriate to compare these models to the version of Saturn's transmission spectrum that did not include refraction (see \S\ref{sec:no_ref}) than to the Saturn's actual transmission spectrum (Fig. \ref{fig:tts_transit_all}, top panel). To first-order, Saturn's transmission spectrum and the spectrum from the self-consistent atmosphere model showed good agreement. Yet, at various locations in the spectrum (i.e. 1.49 $\mu$m, 1.96 $\mu$m, and 2.93 $\mu$m), the atmosphere model exhibited opacity where the transmission spectrum of Saturn did not. These mismatches were due to the existence of gaseous ammonia at low pressures in the self-consistent model, which is not found in Saturn. The chemistry of the model naturally allowed for NH$_3$ condensation and depletion from the gas phase when the temperature-pressure profile became sufficiently cold. However, if the temperature-pressure profile converged to warmer temperatures at low pressures (a warm stratosphere), then the model included a reappearance of gaseous NH$_3$ at low pressure. In reality, Saturn's atmosphere acts like a cold trap, condensing most of the NH$_3$ into a cloud layer near the two-bar pressure level. 

Therefore, we created a second model where all parameters were kept the same but the  gaseous NH$_3$ abundance was forced to zero in the transmission spectrum calculation. This ammonia-free model yielded a substantially better fit to Saturn's transmission spectrum, although some inconsistencies remained: 
\begin{itemize}
\item At 1.27 $\mu$m, 1.58 $\mu$m, 2.08 $\mu$m, 2.96 $\mu$m, and beyond 4.20 $\mu$m the ammonia-free model decreased to values of $\delta_{\lambda}$ below the critical depth set by refraction (see \S\ref{sec:no_ref}) and even below the presumed location of the NH$_3$ cloud deck. In reality, rays could not probe these depths during mid-transit.
\item The self-consistent models displayed continuum absorption due to scattering by aerosols and H$_2$ at wavelength shorter than 1.6 $\mu$m. Although haze is present in Saturn's atmosphere, it was not detected in the transmission spectrum.
\item The glaring disagreement near 3.4 $\mu$m resulted from gaseous ethane and an asymmetric C-H stretching mode of an unknown aliphatic hydrocarbon chain (\S\ref{sec:tts}).
\item Saturn's transmission spectrum displayed opacity near 3.76 $\mu$m that the was not reproduced by the self-consistent atmosphere models. This feature may have been due to gaseous ethane \citep{Sharpe2004,Maltagliati2015}.
\item The peaks of the methane features at 1.15, 1.38, and 2.30 $\mu$m were underestimated by the self-consistent atmosphere models. This may have resulted from errors in either the methane band \Cassini data or the line-by-line opacities of methane used in the self-consistent atmosphere models \citep{Freedman2008}.  Regarding the latter, recent updates to the \textit{ExoMol} database \citep{Yurchenko2014} could potentially explain the observed discrepancies. However, the \citet{Yurchenko2014} results primarily explored the opacities of methane at high temperatures, up to 1,500 K. We would not expect these new line lists to be more appropriate for a model of Saturn's atmosphere (at $\sim$140 K) than the \citet{Freedman2008} results, which specifically apply to cold atmospheres. Other explanations for the discrepancies in the methane feature peaks include opacity from other unidentified species or a disequilibrium process occurring in the region sampled by the observation. Photochemical models and observations suggest that methane destruction occurs near the micro-bar level in Saturn's atmosphere \citep{Moses2005,Fouchet2009}. Production of methane deeper in Saturn's atmosphere to replenish loss due to photolysis may explain the observed excess.
\end{itemize}

\subsection{Refraction and the Transmission Spectrum}\label{sec:no_ref}

Atmospheric refraction determined the minimum altitude rays could probe during mid-transit and therefore the minimum value of transit depth in the transmission spectrum. Consequently, the transmission spectrum did not contain information about the structure or composition of the atmosphere below the critical altitude. We recalculated the transmission spectrum forcing $\nu_{\lambda}=$ 0 at all wavelengths in order to determine what features, if any, were blocked by refraction. In this scenario, rays traveled in straight lines through Saturn's atmosphere and the decrease in flux was entirely due to absorption ($\sigma_{\lambda}$). The resulting transmission spectrum is shown in the bottom panel of Fig. \ref{fig:tts_transit_all}.

The methane features in this ``$\nu_{\lambda}=0$'' transmission spectrum were nearly identical to those in the original transmission spectrum. This was not surprising since refraction effects were minimal in those portions of the spectrum. Away from the methane features, however, rays probed deeper altitudes in Saturn's atmosphere revealing several features that were not present in the original spectrum. First, from comparison to the one-dimensional atmosphere model we found empirical evidence for not being able to probe deeper than approximately two bars, which appeared to be due to a gray opacity source across all wavelengths, presumably the top of the NH$_3$ cloud layer. Second, the minimum depth near 2 $\mu$m did not appear to be set by the same feature that limited the rest of the spectrum. Instead, the opacity at 2 $\mu$m was likely due to C$_2$H$_2$ absorption.

Since rays that experienced the greatest deflection in Saturn's atmosphere originated near the solar limb, the inclusion of limb-darkening could reduce the effects of refraction on the transmission spectrum. As shown by Eq. \ref{eq:area}, including limb-darkening would result in lower $T_{\lambda}$-values thereby increasing $A_{\rm eff, \lambda}$. Consequently, the continuum level of the transmission spectrum may reside slightly above the critical depth set purely by atmosphere refraction. This effect would be negligible for most of Saturn's near-infrared spectrum where the variation in intensity across the solar disk is minimal. However, limb-darkening could not be neglected at shorter wavelengths and could alter the optical transmission spectra of planets with highly refractive atmospheres. For composite transmission spectra that span multiple regimes of the electromagnetic spectrum, special care must be taken to account for stellar limb-darkening.

Calculating Saturn's transmission spectrum with $\nu_{\lambda}=0$ revealed that refraction can suppress features in transmission spectra. This result has been discussed in several previous studies involving refraction and transmission spectroscopy \citep[i.e.][]{Sidis2010,GarciaMunoz2012,Misra2014,Betremieux2013,Betremieux2014,Betremieux2015a}. Although the effects of refraction have been largely unimportant in previous observations of hot giant exoplanet atmospheres \citep[e.g.][]{Hubbard2001}, our results suggest that refraction may be critical to future investigations of giant, long-period exoplanet atmospheres.

\section{Discussion}\label{sec:disc}

\subsection{Implications for Exoplanet Atmosphere Models}

Typical models from the exoplanet atmosphere context reproduce most of the major features, due to methane absorption, across the entire wavelength range.  However, the single largest absorption feature, likely due to gaseous ethane and an unknown aliphatic hydrocarbon derived from methane-based photochemistry \citep{Atreya2005}, was absent from the model. Having opacity between 3.3 and 3.5 $\mu$m, this large feature is particularly alarming because it could influence the transit depth of an exoplanet observed in Channel 1 of the Infrared Array Camera on the \textit{Spitzer Space Telescope}, which is centered at 3.6 $\mu$m \citep{Fazio2004}. This suggests that exoplanet atmospheres at all temperatures may harbor surprises that cannot be easily diagnosed with broad-band photometry.

Minor disagreements between the self-consistent models and the transmission spectrum such as the peak-to-peak sizes of the methane features are also troubling. These mismatches may be caused by local disequilibrium processes (e.g. temperature variations, zonal winds) that are difficult to predict and model. As observations of exoplanet atmospheres progress to ever-greater precision, second-order effects such as these will become increasingly important.

\subsection{Clouds and Transit Transmission Spectra}

Although clouds are present in Saturn's atmosphere at nearly every latitude \citep{Baines2005}, Saturn's transmission spectrum is not flat to 90 ppm. Furthermore, the lowest depth a ray can probe at mid-transit is determined by refraction and not clouds.\footnote{We note that the rays could likely reach the cloud deck at times before and after mid-transit \citep[e.g.][]{Misra2014}.} Therefore, the role of clouds in the transmission spectra of cold, long-period exoplanets may not be as restrictive as that of clouds in warm Earth- and mini-Neptune-sized exoplanets. It is, of course, possible that this solar occultation only probed a relatively cloud-free portion of Saturn's atmosphere. However, variability in Saturn's cloud structure is expected to develop gradually and over large ranges of latitude and longitude \citep{PerezHoyos2006}, making it unlikely that these observations were unique to a specific time or location.

\subsection{Transmission Spectroscopy of Cold Gas Giants}

Saturn's transmission spectrum displays molecular absorption features on the order of 90 ppm, suggesting that transmission spectroscopy is a viable technique to study the atmospheres of cold giant exoplanets. Cold atmospheres can be hosted by planets with extremely long orbital periods (such as Saturn) or by those on shorter orbits around cooler stars. Despite their rarity, giant planets orbiting later-type stars represent an accessible starting point for studies of cold giant-planet atmospheres outside of the solar system. Of all the known \emph{transiting} exoplanets, very few are expected to have cold atmospheres with methane-dominated chemistry akin to Saturn. The best candidate may be Kepler-421b, a Uranus-sized exoplanet orbiting a G9 dwarf star with a period of $\sim$704 days \citep{Kipping2014a}. Assuming a Uranian albedo, the equilibrium temperature of Kepler-421b would be $\sim$185 K. Although the mass of this planet is unknown, its supposed formation location within its protostellar disk suggests that it is likely to be an icy gas giant versus a rocky planet with a gaseous envelope \citep{Kipping2014a}.

A full investigation of the detectability of molecular features in Kepler-421b's atmosphere is beyond the scope of this work. However, if the atmospheric chemistry of Kepler-421b is similar to that of Jupiter or Saturn, we might expect to see substantial methane features in the transmission spectrum. Such a detection would benefit theories of planet formation and migration and would also be the first identification of an active methane cycle occurring in an exoplanet atmosphere.

Additional giant exoplanets with cold atmospheres may be discovered in the near future. Based on expected yields from \citet{Sullivan2015}, the \textit{Transiting Exoplanet Survey Satellite} \citep[\textit{TESS},][]{Ricker2015} is expected to find around a half-dozen giant planets with radii of 6 to 22 R$_{\earth}$ and periods of several hundred days. The cold atmospheres of these potential planets, in addition to that of Kepler-421b, could be probed with follow-up observations by the \textit{Hubble Space Telescope} (\textit{HST}) or the \textit{James Webb Space Telescope} (\textit{JWST}).

\subsection{Observing a Jupiter- or Saturn-Twin Exoplanet}

Cold exoplanets with orbital periods of several hundred days represent a waypoint on the path towards detecting and characterizing giant planets analogous to those in our solar system. However, the challenges associated with observing a long-period Jupiter- or Saturn-twin exoplanet in transit necessitate a different approach than has been previously applied to short-period exoplanets. In the following sections we assess the feasibility of a survey to detect and characterize Jupiter- and Saturn-twin exoplanets in the near future.

\subsubsection{Detectability and Occurrence}\label{sec:occur}

The \textit{a priori} probability of observing a cold, long period exoplanet in transit can be estimated by multiplying the geometric transit probability by the planet occurrence rate. For the purposes of this calculation, we consider exoplanets with periods 4.33\e{3} days $ < P < $ 1.08\e{4} days and masses $0.3 M_J < M < 10 M_J$, where $M_J$ is the mass of Jupiter and 0.3$M_J$ is the mass of Saturn. This period range extends from that of Jupiter ($\sim$11.9 years) to that of Saturn ($\sim$29.5 years). We assume all observations achieve a high enough signal-to-noise ratio to detect 100\% of the geometrically observable transits, which cause decrements in flux on the order of 1\%.

The geometric transit probability for a circular orbit is the inverse of the planet's orbital distance divided by the radius of the host star: $(a/R_{\star})^{-1}$. If we consider Sun-like host stars with planets in circular orbits\footnote{The orbits of Jupiter and Saturn have eccentricities of 0.0489 and 0.0565, respectively. For the purpose of this calculation, we assume simple circular orbits versus accounting for a distribution of eccentricities \citep[e.g.][]{Kipping2013,Kipping2014b}.} with periods in the aforementioned range, the geometric transit probability ranges from $\sim$0.05 to $\sim$0.09\%.  

Occurrence rates of Jupiter- and Saturn-like exoplanets are difficult to estimate because previous transit and radial-velocity surveys are not complete to the long periods associated with these planets. However, direct-imaging observations have suggested that the occurrences derived from radial-velocity surveys can be extrapolated to describe planets at orbital distances up to 100 AU \citep{Brandt2014}. With this in mind, we estimate the occurrence rate of Jupiter- and Saturn-like exoplanets assuming that the probability ($dp$) of a star hosting a planet with mass spanning [$M$, $M+dM$] and orbital period spanning [$P$, $P+dp$] is

\begin{equation}\label{eq:occur}
dp = C \left (\frac{M}{M_0} \right )^{-\alpha} \left (\frac{P}{P_0} \right )^{-\beta} \frac{dM}{M}\frac{dP}{P}
\end{equation}

\noindent where $C$, $\alpha$, and $\beta$ are constants and $M_0$ and $P_0$ are fiducial values chosen to be 1 $M_J$ and 1 day, respectively \citep{Tabachnik2002}. We adopt the values $C = 1.04\e{-3}$, $\alpha = 0.31 \pm 0.2$, and $\beta = -0.26 \pm 0.1$ for FGK dwarf stars from \citet{Cumming2008}, a radial-velocity survey complete in the ranges 2 days $< P < $ 2000 days and $M \ge 0.3M_J$. As shown by \citet{Kipping2014a}, the \citet{Cumming2008} distribution strongly agrees with the observed occurrences rates in the \Kepler sample \citep{Fressin2013}. We integrate Eq. \ref{eq:occur} over the ranges [0.3$M_J$, 10$M_J$] and [4.33\e{3} days, 1.08\e{4} days] to find an occurrence rate of $\sim$2.91\%. Therefore, the \textit{a priori} probability of observing a long-period, giant exoplanet in transit around an FGK dwarf star ranges from $\sim1.5\e{-3}$ to $\sim2.6\e{-3}$ percent.

Based on these probabilities, we estimate that $\sim$38,500 stars would have to monitored for 11.9 years in order to find a single Jupiter-analogue, or $\sim$66,700 stars for 29.5 years for a single Saturn-analogue. Clearly, any survey to find long-period, giant exoplanets in transit must observe a large ($>$10$^5$) number of stars to make a detection on a practical time scale.

\subsubsection{Survey for Long-Period, Giant Exoplanets}\label{sec:survey}

\begin{figure}
\centering
\includegraphics[scale=0.38]{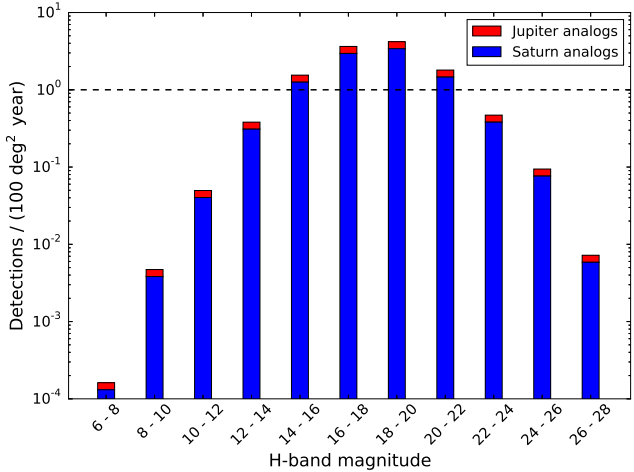}
\caption{Expected number of long-period, giant planet transit detections per 100 deg$^2$ after a single year of observation. We estimate these detection rates using a synthetic catalog of FGK dwarf stars generated with the TRILEGAL simulation code \citep{Girardi2005} and the \textit{a priori} transit probabilities from \S\ref{sec:occur}. A horizontal line is drawn at unity for reference. For a Saturn-analog exoplanet, the Wide Field Camera 3 (WFC3) aboard the \textit{Hubble Space Telescope} (HST) could make a 5$\sigma$ detection of transmission spectrum features for stars with H-band magnitudes $m_H \lesssim 9.2$. If the rates in this figure are extrapolated to cover the whole sky, we would expect one detection per year suitable for characterization with \textit{HST}-WFC3.}
\label{fig:survey}
\end{figure}

We estimate the number of long-period (4.33\e{3} days $<P<$ 1.08\e{4} days), giant ($0.3 M_J<M<10 M_J$) exoplanet detections around FGK dwarf stars using a stellar population generated by the Tridimensional Model of the Galaxy\footnote{\url{http://stev.oapd.inaf.it/cgi-bin/trilegal}} synthesis code \citep[TRILEGAL,][]{Girardi2005}. Using the default input parameters, we generate a stellar population in a 10 deg$^2$ field centered on the galactic coordinates of the \Kepler field ($l=76^{\circ}$, $b=$ +14$^{\circ}$) with limiting H-band magnitude $m_H<32$. Of the full sample ($\sim$3.3\e{6} stars), we only consider stars with effective temperatures and luminosities in the fiducial ranges 3,800 K $<T_{\rm eff}<$ 7,000 K and $-1.5 <log_{10}(L/L_{\odot})< 1.0$ (where $L_{\odot}$ is the solar luminosity) in an attempt to limit the sample to FGK dwarf stars. The stars are grouped into bins of width two magnitudes between $m_H=6$ and $m_H=28$; bins on either side of these limits contain zero stars. The total star counts in each bin are multiplied by the \textit{a priori} probabilities calculated in \S\ref{sec:occur} and divided by the periods of Jupiter and Saturn to determine the final detection rates for long-period, giant exoplanets (Fig. \ref{fig:survey}). Note that we present detections per 100 deg$^2$ per year; TRILEGAL limits the field area to 10 deg$^2$ so we simply increased the number of stars in the sample by a factor of 10. 

It is important to once again note that we assume this type of survey would recover 100\% of the observable transits. For reference, the transits of Jupiter and Saturn across the Sun would cause decrements in flux of $\sim$1.1\% and $\sim$0.7\%, respectively. To observe transits of this magnitude, this survey would not require a space-based telescope. Instead, arrays of ground-based, robotic telescopes, akin to MEarth \citep{Irwin2009} or MINERVA \citep{Swift2015} could be used to detect transit events. An all-sky, array telescope such as the ``Evryscope'' would also be a highly appropriate instrument for this type of survey \citep{Law2015}; the construction of the Evryscope is in part motivated by the ability to observe giant planets transiting nearby bright stars. Further characterization of these cold giant exoplanets would, however, require more powerful observing facilities.

\subsubsection{Target-of-Opportunity Follow-Up Observations}\label{sec:ToO}

The long periods of these exoplanets necessitate immediate follow-up characterization. Fortunately, long periods also result in long transit durations. From the point-of-view of a distant observer, the transits of Jupiter and Saturn across the solar disk would last $\sim$23 and $\sim$57 hours, respectively.\footnote{Assuming circular orbits with inclinations of 90$^{\circ}$.} These transit durations are long enough such that target-of-opportunity campaigns with facilities such as \textit{HST} or \textit{Spitzer Space Telescope} could be activated in time to characterize the exoplanet's atmosphere. The infrastructure for this type of observing program is already in place in the field of gamma ray bursts (GRBs). Since 2004, the \textit{Swift} Mission has been observing GRBs and relaying the coordinates and data to the GRB community worldwide in just a matter of seconds \citep{Gehrels2004}.

To demonstrate the ability of current facilities to characterize the atmospheres of cold giant exoplanets, we specifically consider the case of a Saturn-Sun analog observed with the Wide Field Camera 3 (WFC3) aboard \textit{HST} and the Infrared Array Camera (IRAC) aboard \textit{Spitzer}. Upon observing the slow ($\sim$3.3-hour), deep ingress, the survey telescopes would issue an alert calling for the activation of the target-of-opportunity programs. Under ideal conditions, \HST could begin observing the transit 24 hours after activation,\footnote{\url{http://www.stsci.edu/hst/HST_overview/documents/uir/ToO-UIR.pdf}} capturing the final $\sim$29 hours of the transit. \Spitzer normally requires 48 hours to initiate a target-of-opportunity program,\footnote{\url{http://ssc.spitzer.caltech.edu/warmmission/ddttoo/whattoo/}} which leaves an insufficient amount of time to characterize the planet's atmosphere. For the purposes of this thought-experiment, however, we will consider \textit{Spitzer's} ability to characterize a Saturn-twin exoplanet atmosphere regardless of the 48-hour turnaround time. 

\begin{figure*}
\centering
\includegraphics[scale=0.64]{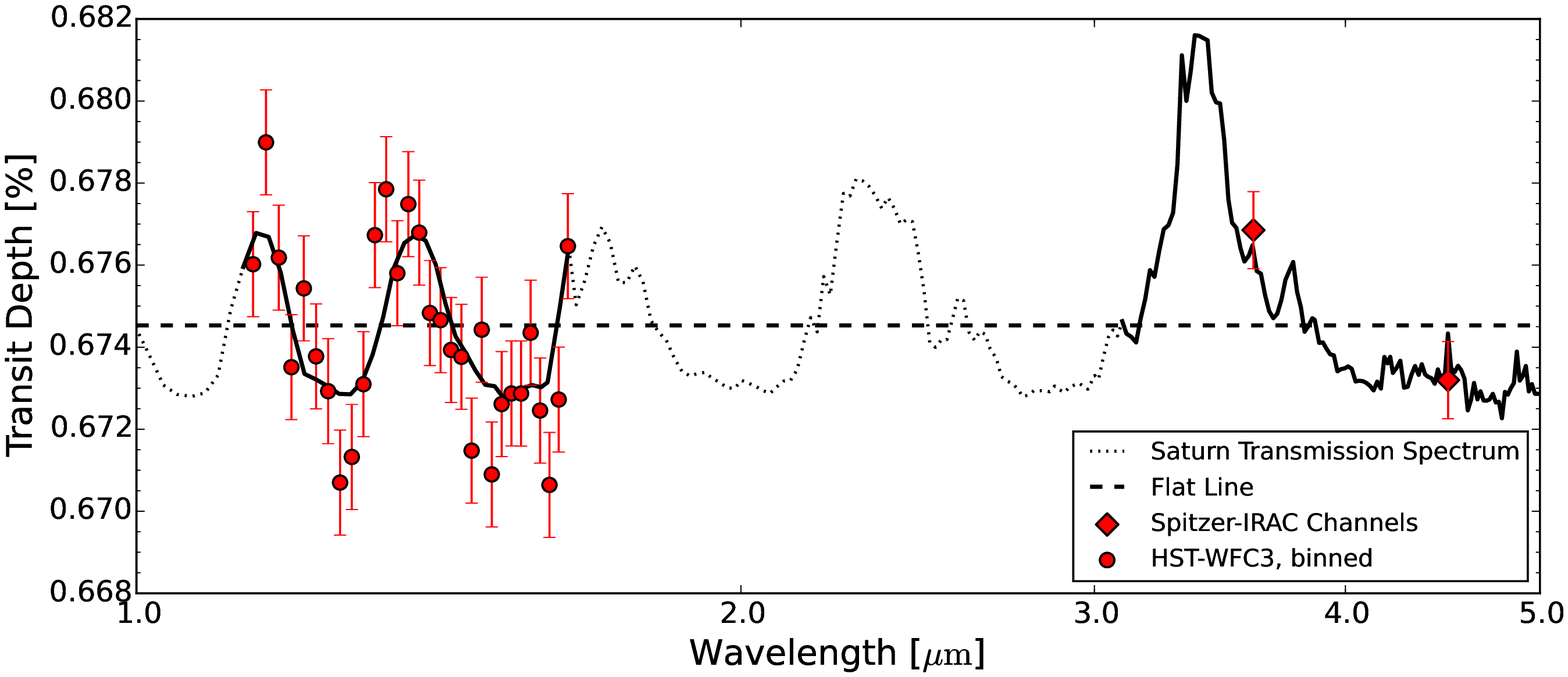}
\caption{Simulated \textit{HST}-WFC3 and \textit{Spitzer}-IRAC observations of Saturn's transmission spectrum. The spectrum is solid in regions sampled by either \HST or \Spitzer and dotted elsewhere. The red circles are the expected \textit{HST}-WFC3 data points, binned by 2 resolution elements and scattered with random Gaussian noise of $\sim$13 ppm. The red diamonds are the expected \textit{Spitzer}-IRAC data points with an uncertainty of $\sim$9 ppm. Note that \Spitzer cannot observe both channels simultaneously. In each case, the quoted precision is the requirement to distinguish the features in the transmission spectrum from a flat line (black, dashed line) to 5$\sigma$ confidence. To achieve this precision in 0.5 transits of a Saturn-twin exoplanet across a solar-type star (see text), \HST would be limited to stars with H-band magnitudes $m_H \lesssim$ 9.2. \Spitzer could achieve the displayed precision in a single channel (either 3.6 or 4.5 $\mu$m) for stars with $m_H \lesssim$ 3.4 but would require additional observing time to observe the the other channel. Therefore, \Spitzer observations of this type of target are infeasible.}
\label{fig:hst}
\end{figure*}

In Fig. \ref{fig:hst}, we show the expected \textit{HST}-WFC3 and \textit{Spitzer}-IRAC transmission spectrum of a Saturn-analog exoplanet derived from the \textit{Cassini}-VIMS transmission spectrum. The spectral resolution of \textit{HST}-WFC3 nearly matches that of \textit{Cassini}-VIMS in this wavelength range, so a convolution to match resolution is unnecessary.  We bin the high signal-to-noise ratio portion of the \textit{HST}-WFC3 spectrum by 2 resolution elements yielding 30 data points between 1.13 to 1.65 $\mu$m. Then, by repeatedly scattering the data points with random Gaussian noise of 1 to 50 ppm, we determine that a minimum precision of 12.8 ppm is required to distinguish features in the \textit{HST}-WFC3 spectrum from a flat line to 5$\sigma$ confidence. For the simulated \Spitzer observations, we integrate the IRAC bandpasses over the transmission spectrum of Saturn to determine the 3.6 and 4.5 $\mu$m data points. We estimate that an uncertainty of 9.4 ppm is required in each \Spitzer data point to rule out a flat spectrum to 5$\sigma$. It is critical to note that \Spitzer cannot observe both IRAC channels simultaneously. Each data point must be obtained individually.

These uncertainties set upper limits on the magnitudes of Saturn-hosting stars that are amenable to characterization with \textit{HST}-WFC3 and \textit{Spitzer}-IRAC. To determine this limit for \HST, we consider a large variety of observing strategies (e.g. staring versus spacial scan modes, various slew rates \citep{McCullough2012}, subarray sizes, and readout configurations) over a range of H-band magnitudes matching the output from the TRILEGAL simulation ($6 < m_H < 28$). In each case, we assume \HST observes the transiting system for 36 consecutive orbits: 18 during the final half of transit, and 18 out-of-transit orbits to establish a precise baseline for the stellar flux. We assume the host star is visible for 56 minutes of the 96-minute orbit before Earth occultation, similar to the stars in the \Kepler field. The nominal exposure time is set by the chosen readout configuration, and the corresponding signal-to-noise ratio per resolution element per exposure is estimated using the \textit{HST}-WFC3 exposure time calculator.\footnote{\url{http://etc.stsci.edu/etc/input/wfc3ir/spectroscopic/}} For each observing configuration, we use the Phase II Astronomer's Proposal Tool\footnote{\url{http://www.stsci.edu/hst/proposing/apt/}} Orbit Planner to determine the number of exposures we can obtain per \HST orbit and make a final estimate of the precision of the transmission spectrum. 

The result of this calculation is that with only a half transit, \textit{HST}-WFC3 can make a 5$\sigma$ detection of atmospheric features in the transmission spectrum of a Saturn-analog if the host star has $m_H \lesssim 9.2$. As displayed in Fig. \ref{fig:survey}, fewer than one Saturn-analog detection is expected per 100 deg$^2$ per year. However, if we extrapolate these detection rates to cover the entire sky, we would expect approximately one detection per year amenable to characterization with \textit{HST}-WFC3.

Considering that \Spitzer cannot respond quickly enough to characterize a transiting Saturn-analog exoplanet and each channel must be observed individually, we estimate its limiting host-star magnitude in less detail than for \textit{HST}. If we assume photon-limited observations, we can loosely estimate uncertainties by scaling those obtained for a previous \textit{Spitzer}-IRAC observation of a solar-type star. For 55 Cancri ($m_H$=4.14), \citet{Demory2011} achieved 63-ppm-precision in IRAC's 4.5 $\mu$m band over 4.97 hours of observation. If \textit{Spitzer}-IRAC could only observe 0.5 transits (29 hours in transit + 29 hours out of transit) of a Saturn-twin exoplanet orbiting a solar-type star \emph{in a single channel}, then the 9.4 ppm precision requirement would limit the host star H-band magnitudes to $m_H \lesssim 3.4$. \Spitzer would then have to wait until the following transit event to obtain observations in the other channel. This first-order approximation demonstrates that \HST is by far the most appropriate currently-operational instrument for characterizing the atmospheres of cold, long-period exoplanets.

The success of this hypothetical survey is contingent upon the ability of the survey telescopes to quickly and accurately identify long-period, giant exoplanet transits. This would require immediate, automatic data reduction and analysis. For stars brighter than $m_H \approx$ 9.2 and transit durations longer than $\sim$57 hours, there is some flexibility that would allow for human intervention. Still, distinguishing false-positives from actual events would be a major challenge to this approach. As in any other wide-field transit survey, false alarms may result from variations in instrument sensitivity, weather, or other astrophysical sources such as stellar variability or unknown stellar companions. To the extent that it is possible, explicit target selection and advance ``snapshot'' observations could limit astrophysical false positives. To reduce the false positives due to eclipsing binary stars, the target-of-opportunity program could also involve obtaining a spectrum of the target in search of two sets of spectral lines.

\section{Concluding Remarks}

Studies of solar system analogs provide a useful method of ``ground-truthing'' the techniques and models frequently applied to exoplanets. Exhausting the resources provided by decades of work in the planetary sciences will greatly aid the burgeoning field of exoplanetary science. The usefulness of missions such as \Cassini and Juno, which is currently \textit{en route} to Jupiter, extends beyond the solar system to the cold, long-period regime of exoplanets. 

The \Kepler Mission has discovered a great variety of Earth-sized exoplanets. Future efforts to discover and characterize cold Jupiters and Saturns may find that a similar diversity exists among giant gaseous planets. These efforts will put the giant members of our solar system in a greater context, thereby allowing for a better understanding of the formation and evolution of our entire solar system.

\acknowledgements

We wish to thank the anonymous referee for thoughtful suggestions that improved this work. We also wish to thank Brandon Harrison, Andrew Vanderburg, Bryce Croll, and the \textit{Cassini}-VIMS team. P.A.D. gratefully acknowledges support from the Massachusetts Space Grant Consortium. This paper includes data collected by the \Cassini Mission, funding for which was provided by NASA and ESA. This research made use of the Exoplanet Orbit Database and the Exoplanet Data Explorer at \url{exoplanets.org} and the NASA Exoplanet Archive, which is operated by the California Institute of Technology, under contract with NASA under the Exoplanet Exploration Program. This research made use of the USGS Integrated Software for Imagers and Spectrometers. This research also made use of the Massachusetts Green High Performance Computing Center.

{\it Facilities:} \facility{Cassini (VIMS)}

\bibliographystyle{apj}
\bibliography{references}

\end{document}